\documentclass[fleqn,10pt]{wlscirep}
\usepackage[utf8]{inputenc}
\usepackage[T1]{fontenc}
\title{Signal Amplification in a Time-Modulated Transmission Line and the Loss Effect}

\author[1,2,*]{Mohamed F. Hagag}
\author[3]{Thomas R. Jones}
\author[2]{Karim Seddik}
\author[3]{Dimitrios Peroulis}
\affil[1]{Electronic Engineering Department$,$ Military Technical College$,$ Cairo  11766$,$ Egypt}
\affil[2]{Electronics and Communications Engineering Department$,$  American  University   in  Cairo$,$  Cairo   11835$,$  Egypt}
\affil[3]{Elmore Family School of Electrical and Computer Engineering$,$ Purdue University$,$ West  Lafayette$,$  Indiana   47907$,$ USA}
\affil[*]{corresponding.m.a.fouadhagag@ieee.org}

\begin{abstract}
We investigate and simulate signal amplification in a transmission line (TL) with time-modulated characteristic impedance $Z_o$. Periodically varying $Z_o$ is achieved by loading TL with a sinusoidally time-modulated capacitor (TMC). For a detailed study, three models are considered: a lossless L-C TL lumped model loaded with shunt infinite quality factor (Q) TMC, a TL loaded with a shunt infinite Q TMC, and finite Q TMC. By solving the eigenvalue problem in all models, dispersion diagrams (DD) are plotted with a created momentum band gap (MBG) at a modulation frequency double the signal frequency. Within MBG, only imaginary frequencies are found and correlated to MBG width and signal growth level. Using Harmonics Balance (HB) and Transient Simulation (TS), signal amplification is confirmed, and the obtained results are consistent with the DD outcomes. In the second model, the effect of TL length on amplification is investigated and explained by studying the unit cell's Bloch impedance ($Z_{Bloch}$). The loss effect is considered by adding a series resistance (Rc) to the third model's TMC (finite Q). Decreasing amplification levels, confirmed by circuit modeling, due to the increase of Rc value is explained by studying real and imaginary DDs and the attenuation constant. 
\end{abstract}
\begin{document}

\flushbottom
\maketitle
%\thispagestyle{empty}

%\noindent Please note: Abbreviations should be introduced at the first mention in the main text – no abbreviations lists. Suggested structure of main text (not enforced) is provided below.

\section{Introduction}
	Temporal modulated media is an incoming technique that unleashes enormous opportunities for artificial electromagnetic media \cite{10.1117/1.AP.4.1.014002,doi:10.1126/science.aat3100,Engheta+2021+639+642}. It is realized by varying the media characteristics periodically with time following a specific modulation waveform. The interest in engineering artificial media and materials with time-modulated properties has recently increased. One attractive proposed property of time-modulated media is magnetless non-reciprocity \cite{Yu2009,Ruesink2016}. Many magnetless non-reciprocal components are presented in the literature, such as filters, circulators, and power dividers \cite{9769703,9224096,8058574,8439077,ZANG2021153609}. The non-reciprocity is created when a sub-component property (unit cell) is time modulated with a speed lower than the signal speed with sufficient successive phase shift, creating a spatiotemporal modulated component \cite{PhysRevB.92.100304,7304706}. Other interesting applications are proposed, such as creating an effective magnetic field for photons \cite{Fang2012} and synthetic dimensions \cite{Lustig18,Wang2021}.  
	
	More interesting phenomena are discovered when modulating media properties in time at high speed. This was first proposed as a photonic time crystal (PTC) or generally time crystal (TC) \cite{Sharabi:22,Zeng2017}. A photonic crystal (PC) is a periodic structure at which the signal wave number (K) is engineered through space to create a frequency band gap (FBG); within the FBG, the signal is forbidden to propagate\cite{Joannopoulos2008-mv,Sievenpiper1999HighimpedanceES}. As an analogy to PCs, TCs are created when periodic structure properties are modulated in time and kept uniform in space. With a sufficient time modulation speed, band gaps in momentum (MBG) are created inside which a PTC possesses a non-Hermitian nature. As a result, the signal is amplified exponentially in the MBG \cite{Lustig18,10.1117/1.AP.4.1.014002} in contrast to the signal decay in the FBG \cite{Joannopoulos2008-mv,Sievenpiper1999HighimpedanceES}. To realize an MBG in a TC, its properties must be modulated in time with a speed higher than the signal propagation speed inside the TC. Realizing this in the microwave regime is often more attainable \cite{GGGGGG1}. Within this regime, there are adaptive components such as varactors and photo-diodes; the change in their properties may be able to catch up with the modulation signal.

	In this work, lossless transmission lines (TL) loaded with time-modulated capacitors (TMCs), which act as 1-D TCs, are analyzed and discussed as a step in realizing signal amplification in the microwave regime. First, the eigenvalue problem is formulated and solved for a lossless TL-lumped model with a TMC. The dispersion diagram is plotted, and the areas of the MBGs are defined. To investigate the circuit performance within the MBG, S-parameter (SP), Transient (TS), and Harmonic Balance (HB) simulations in Keysight Advanced Design System (ADS) software are performed on 9 unit cells. Amplification is observed, and there is an agreement between the DD and simulation results. Second, a TL loaded with  TMCs is considered. The above analyses are applied, and agreement between the DD and simulation results is achieved. The effect of varying the TL lengths on MBGs is discussed and supported by plotting the unit cell Bloch impedance ($Z_{Bloch}$). Finally, to study the loss effect, a series resistance (Rc) is introduced within the TMC to realize a finite Q TMC. Real and complex DDs with different Rc (Q) values are plotted, and weak interaction between harmonics is observed. The results are confirmed using ADS TS and HB analysis. 
	\section{Dispersion equation of a lossless TL lumped model with time-modulated capacitor}
	
\begin{figure}[!t]
		\centering
		\includegraphics[width=.78\linewidth]{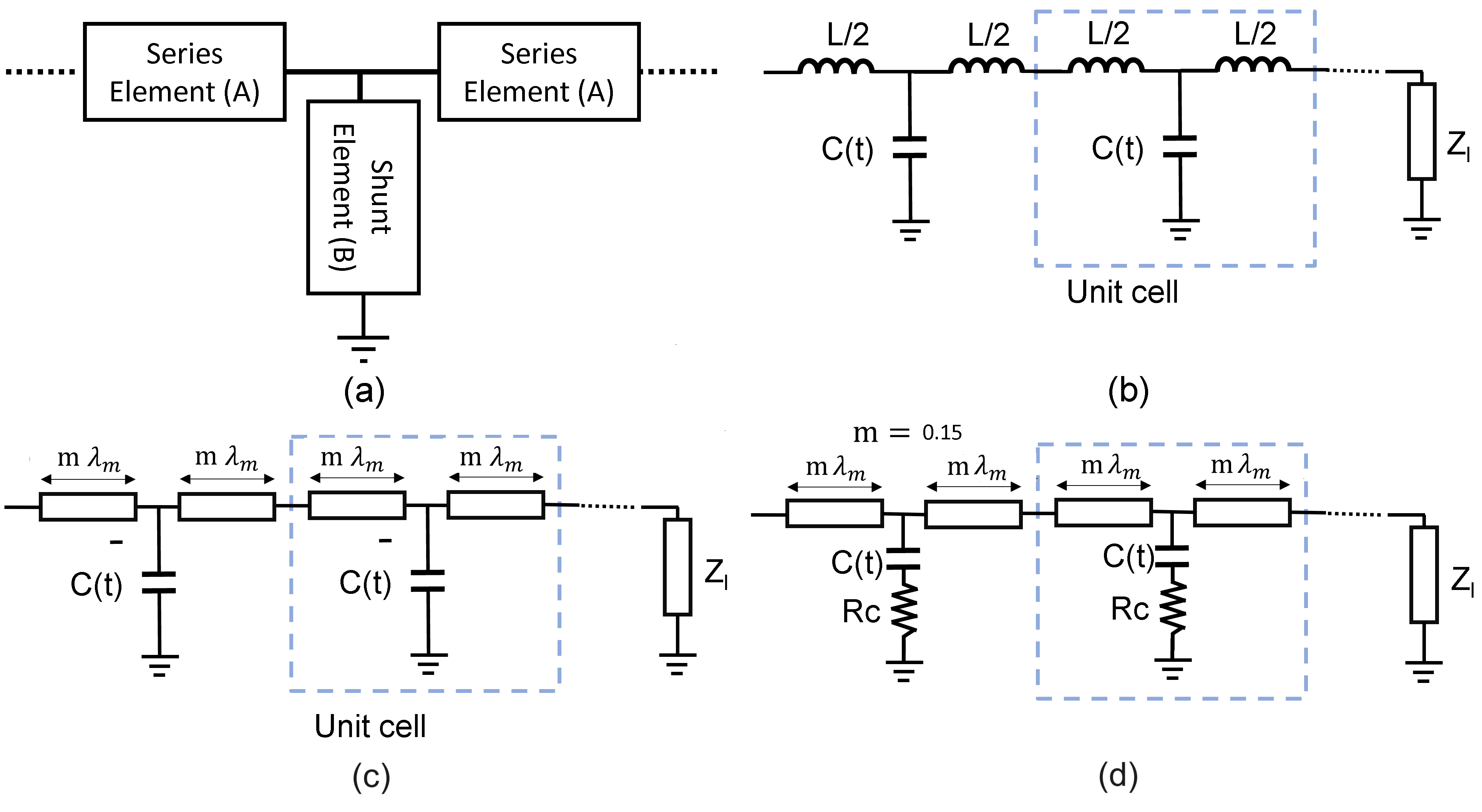}
		\caption{(a) General symmetric unit cell of an ideal transmission line. (b) A unit cell of a lossless lumped model of a transmission line with a time-modulated capacitor. A unit cell of a lossless TL with a length of $m\lambda_m$ and $90~\Omega$ characteristic impedance loaded with (c) infinite quality factor time-modulated capacitor, (d) finite quality factor time-modulated capacitor.}
		\label{MODLE}
	\end{figure}

	This section constructs the dispersion equation of a lossless lumped model TL loaded with a TMC. The unit cell of the loaded TL is shown in Fig.~\ref{MODLE}(a), which is a T-shape symmetric unit cell constructed from two different elements, series, and shunt. The shunt element (B) will always be the time-modulated capacitor. The series element (A) will be a time-invariant inductor. 
	
	\subsection{Transfer matrix of time-modulated capacitor}
	To construct the dispersion equation, transfer matrices of all elements were computed. As in \cite{9063633,Jayathurathnage2021,Elnaggar2021}, a TMC with a period $T$ that follows the function $C(\omega,t)=C(\omega,t+nT)\ $ with $ n\in\mathbb{Z}$ can be expanded into a complex Fourier series as follows
	\begin{equation}
	C\left(\omega,t\right)=\sum_{s=-\infty}^{+\infty}{c_s(\omega)e^{js\omega_Mt}}
	\label{eq:C}
	\end{equation}
	where $c_s(\omega)$ are complex coefficients and $\omega_M$ is the angular modulation frequency. When the TMC is connected to a circuit operating at angular signal frequency $\omega_s$, an infinite number of Floquet harmonics are generated; the voltage or current across the TMC follows the series below \cite{Jayathurathnage2021}
	
	\begin{equation}
	F\left(t\right)=\sum_{n=-N}^{+N}{f_ne^{j\omega_nt}}
	\label{eq:V}
	\end{equation}
	where $f_n$ are the complex coefficients of voltage and current. $N$ is an integer that represents half the number of considered harmonics in the problem and depends on the electrical size of the unit cell. According to \cite{Jayathurathnage2021}, it can be approximated by
	\begin{equation}
	N\ll\frac{\lambda_s}{d}
	\label{eq:N}
	\end{equation}
	where $\lambda_s$ is the signal's wavelength, and $d$ is the length of the unit cell. The harmonic frequencies can be given by
	\begin{equation}
	\omega_n=\omega_s+n\omega_M
	\label{eq:om}
	\end{equation}
	where $\omega_s$ and $\omega_M$ are the main signal and modulation angular frequencies, respectively. Eq. \ref{eq:C} depends mainly on the modulation waveform. In this manuscript, the capacitance is modulated around its nominal value in a sinusoidal manner following
	\begin{equation}
	C\left(\omega,t\right)=C_o(\omega)\left( 1+M_D\cos{(\omega_Mt+\varphi_M}\right))
	\label{eq:C1}
	\end{equation}
	where $M_D$ and $\varphi_M$ are the modulation depth and phase, respectively. $C_o(\omega)$ is the capacitance nominal value. 
	Using (\ref{eq:C1}), (\ref{eq:C}) is reduced to
	
	\begin{equation}
	C\left(\omega,t\right)=C_o(\omega)+C_{-1}(\omega)e^{-j\omega_Mt}+C_1(\omega)e^{j\omega_Mt}
	\label{eq:C3}
	\end{equation}
	where $C_{\pm1}(\omega){=0.5~C_o(\omega) \ M_De}^{\pm j\varphi_M}$.
	
	In general, as discussed in \cite{Jayathurathnage2021,Elnaggar2021}, the TMC generated  current and voltage harmonics are related by 
		\begin{equation}
	i_l=\sum_{n=-\infty}^{\infty}{j~\omega_r~c_{l-n}(\omega_{l})~v_{n}}
	\label{eq:N2}
	\end{equation}
	Consequently, the admittance matrix of the TMC is given by
	\begin{equation}
	\bar{\bar{Y_C}}=j\bar{\bar{W}} P
	\label{eq:YC}
	\end{equation}
	where $\bar{\bar{W}}=diag(\omega_{-N},..,\omega_{-1},\omega_s,\omega_{1},..,\omega_N)$, and $P$ in general is given by
	\begin{equation}
	P=\left(\begin{matrix}\begin{matrix}c_0\left(\omega_{-N}\right)&c_{-1}\left(\omega_{1-N}\right)\\c_1\left(\omega_{-N}\right)&c_0\left(\omega_{1-N}\right)\\\end{matrix}&\cdots&\begin{matrix}c_{-2N}\left(\omega_N\right)\\c_{1-2N}\left(\omega_N\right)\\\end{matrix}\\\vdots&\ddots&\vdots\\\begin{matrix}c_{2N}\left(\omega_{-N}\right)\ &c_{2N-1}\left(\omega_{1-N}\right)\\\end{matrix}&\cdots&c_0\left(\omega_N\right)\\\end{matrix}\right)
	\label{eq:YCC}
	\end{equation}
	$P$ is $(2N+1) \times (2N+1)$ in size. Using a TMC that follows (\ref{eq:C3}), the $P$ matrix is reduced to 
	\begin{equation}
    P=\left(\begin{matrix}\begin{matrix}c_0(\omega_{-N})&c_{-1}(\omega_{1-N})&0\\c_1(\omega_{-N})&c_0(\omega_{1-N})&c_{-1}(\omega_{2-N})\\0&c_1(\omega_{1-N})&c_0(\omega_{2-N})\\\end{matrix}&\begin{matrix}\cdots\\\cdots\\\cdots\\\end{matrix}&\begin{matrix}0\\0\\0\\\end{matrix}\\\begin{matrix}\vdots&\ \ \ \ \ \ \ \ \ \ \ \ \ \ \ \ \vdots&\ \ \ \ \ \ \ \ \ \ \ \ \ \ \ \ \ddots\\\end{matrix}&\ddots&\vdots\\\begin{matrix}0\ \ \ \ \ \ \ \ \ \ \ \ \ \ \ &0\ \ \ \ \ \ \ \ \ \ \ \ \ \ \ \ &0\\\end{matrix}&\cdots&c_0(\omega_N)\\\end{matrix}\right)
		\label{eq:YC1}
  \end{equation}
	
	The transfer matrix of a TMC can be obtained as
	\begin{equation}
	T_C=\left[\begin{matrix}\bar{\bar{Ones}}&\bar{\bar{Zeros}}\\\bar{\bar{Y_C}}&\bar{\bar{Ones}}\\\end{matrix}\right]
	\label{eq:TC}
	\end{equation}
	where $\bar{\bar{Ones}}$ and $\bar{\bar{Zeros}}$ are the unit and zero matrices, receptively. They have the same size of $\bar{\bar{Y_C}}$. This makes the transfer matrix $T_C$ of size $(4N+2) \times (4N+2)$. 

	\begin{figure*}[!t]
		\centering
		\includegraphics[width=1.03\linewidth]{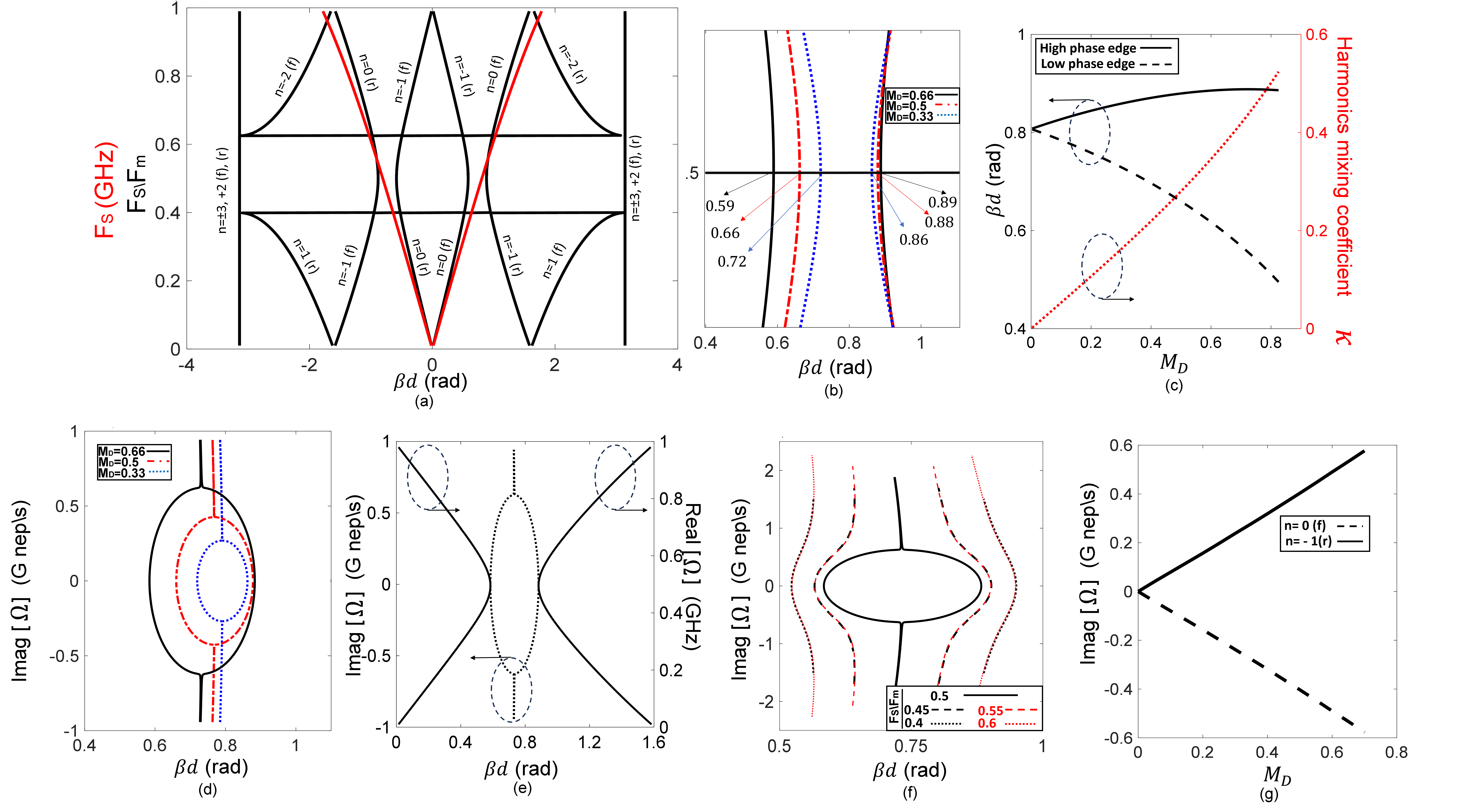}
		\caption{(a) Dispersion diagram of the unit cell in Fig.~\ref{MODLE}(b) with $L= 12.5 $ nH, $C_o =5 $ pF and $M_D=0.66$. In the case of no modulation, the dispersion diagram only reduces to the red curves. (b) Close-up look of the momentum band-gap area at different values of $M_D$. (c) MBG-edge phase variation and harmonics ($\omega_o$ and $\omega_{-1}$)  mixing coefficient variation at different values of $M_D$. 
  %(d) harmonics $\omega_o$ and $\omega_{-1}$ magnitude of eigenvectors components at different values of $M_D$. 
  (d) The imaginary part of complex frequency variation within the MBG at $F_s/F_m=0.5$ for harmonics $\omega_o$ and $\omega_{-1}$. At $M_D=0.66$, (e) Unit cell electrical length variation with the real and imaginary parts (real is fixed at $F_s/F_m=0.5$ for imaginary part variation) of the complex frequency, (f) The imaginary part of complex frequency variation within the MBG at different real part values for harmonics $\omega_o$ and $\omega_{-1}$, (g) For harmonics $\omega_o$ and $\omega_{-1}$, The imaginary part of complex frequency variation with $M_D$ at $F_s/F_m=0.5$ and unit cell electrical length at nominal Capacitance $C_o$ ($\beta d=0.81$).
  %(g) $\beta d$ gradient variation around $F_s/F_m=0.5$ for   harmonics $\omega_o$ and $\omega_{-1}$.
  }
		\label{DDLC}
	\end{figure*}
	\subsection{Dispersion equation}
	In this section, the series element (A) in Fig. \ref{MODLE}(a) is a time-invariant inductor (TII). The impedance matrix of such an inductor in a time-invariant system is given by
	\begin{equation}
	\bar{\bar{Z_L}}=j\bar{\bar{W}} (L/2)
	\label{eq:ZL}
	\end{equation}
	where $L= \bar{\bar{Ones}}\times l_o$, and $l_o$ is the value of the inductance. The transfer matrix of a TII can be obtained by
	\begin{equation}
	T_L=\left[\begin{matrix}\bar{\bar{Ones}}&\bar{\bar{Z_L}}\\\bar{\bar{Zeros}}&\bar{\bar{Ones}}\\\end{matrix}\right]
	\label{eq:TL}
	\end{equation}
	The total transfer matrix of the unit cell can be obtained using
	\begin{equation}
	T_t= T_L \times T_C \times T_L
	\label{eq:TT}
	\end{equation}
	The dispersion diagram can be plotted by first solving 
	\begin{equation}
	T_t X=q X
	\label{eq:dd}
	\end{equation}
	where $X$ and $q$ are the eigenvectors and values, respectively. Once the eigenvalues are obtained, the dispersion diagram is plotted using
	\begin{subequations}
		\begin{equation}
		q=e^{i \gamma d}
		\label{eq:dd11}
		\end{equation}
		\begin{equation}
		\beta d=  real[ -i ln(q)]
		\label{eq:dd1111}
		\end{equation}
		\begin{equation}
		\alpha d=  real[ ln(q)]
		\label{eq:dd111f1}
		\end{equation}
		\label{eq:dd1}
	\end{subequations}
	where $\gamma$ is the propagation constant, $\beta$ is the phase constant, $\alpha$ is the attenuation constant, and $d$ is the length of the unit cell. Using (\ref{eq:dd1},\ref{eq:dd}), plotting the dispersion variation against frequency results in $(4N+2)$ curves; two for each harmonic plus two for the fundamental frequency. This describes each frequency's forward and backward propagation defined in (\ref{eq:om}).

	\begin{figure}[!t]
		\centering
		\includegraphics[width=.65\linewidth]{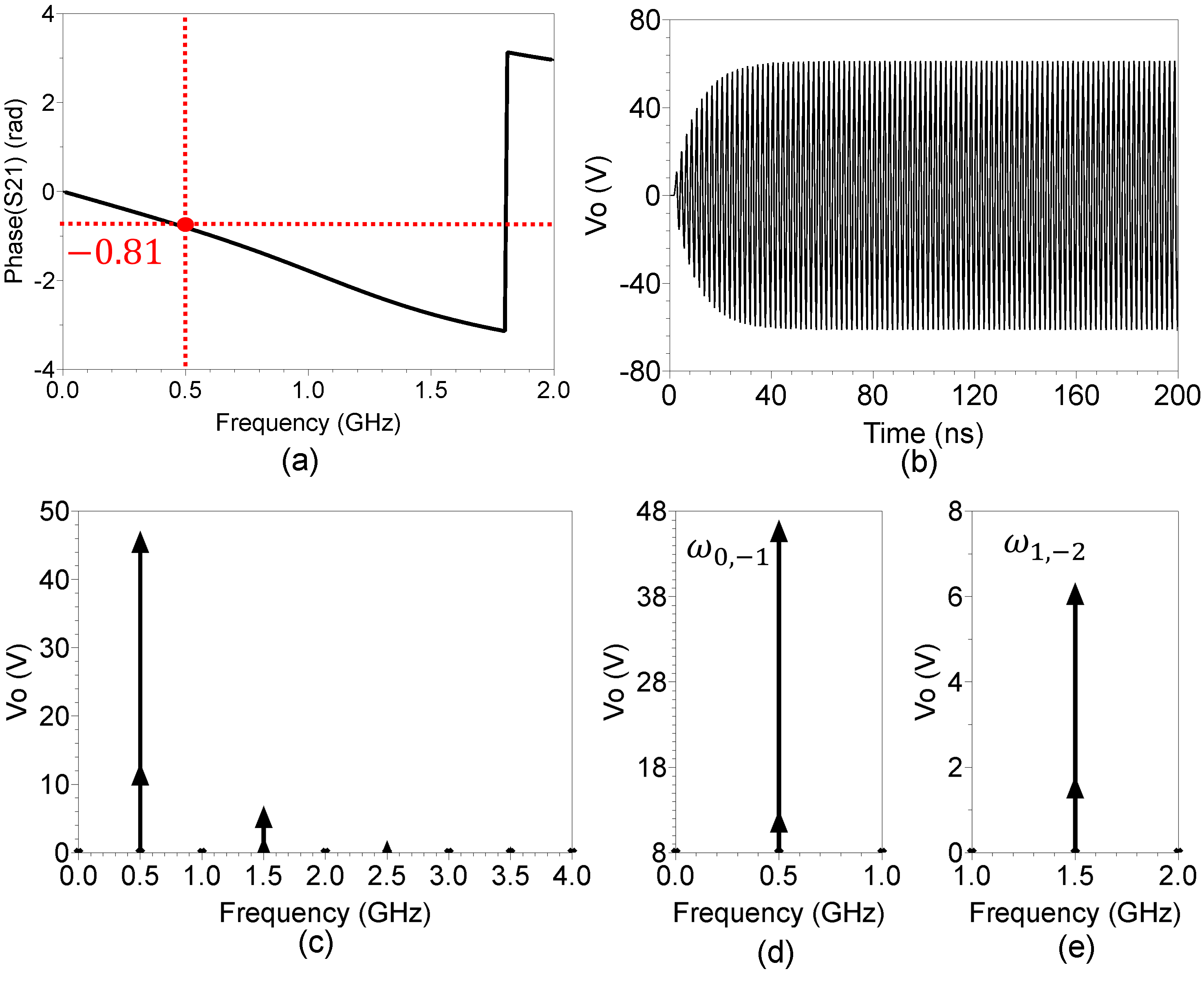}
		\caption{(a) Phase of $S_{21}$ for only one unit cell shown in Fig.~\ref{MODLE}(b) with $L= 12.5$~nH and $C_o =5$~pF (static), S-parameter simulation. For 9 unit cells with 10~V input peak (1 Watt (rms)), TMC at $M_D =0.66$, $L= 12.5$~nH, $C_o =5$~pF, $F_s =0.5$~GHz, $F_m =1$~GHz loaded with a $50~\Omega$ load impedance. (b) Output voltage (TS). (c) Output voltage (HB simulation). (d) Close-up look for the harmonics at 0.5~GHz. (e) Close-up look for the harmonics at 1.5~GHz.}
		\label{lccir}
	\end{figure}

	\section{ Signal Amplification in a lossless TL lumped model with  time-modulated capacitor}
	
	The DD is plotted in Fig.~\ref{DDLC}(a) for the unit cell shown in Fig.~\ref{MODLE}(b). To plot Fig.~\ref{DDLC}(a), numerical values are chosen as follows: $F_m= 1$~GHz, $L= 12.5$~nH, $C_o= 5$~pF, and $M_D= 0.66$. The unit cell has a 50~$\Omega$ characteristic impedance at nominal capacitance $C_o$. Fig.~\ref{DDLC}(a) shows that only the fundamental frequency $\omega_s$ exists in the absence of modulation; the red curves describe the forward and backward propagation dispersion at $\omega_s$ in case of no modulation. With modulation present, black curves describe the dispersion of six considered harmonics (N=3) and the fundamental frequency. All curves are labeled with the harmonic order and type of propagation: forward (f) or backward (r). At $F_s= 0.5$~GHz ($F_m= 2F_s$), strong interaction happens between the fundamental $\omega_s$ and the harmonic $\omega_{-1}$, which operates at the same frequency. Due to the strong coupling between harmonics mentioned above, a region of unstable K-gap is created called MBG \cite{10.1117/1.AP.4.1.014002,wang2023metasurface,sharabi2022spatiotemporal}. At this region, exponential amplification occurs for the signal, and the circuit acts as a parametric amplifier \cite{1125452,Jayathurathnage2021}. However, it is worth mentioning that the DD is not affected by varying the modulation phase $\varphi_M$. As a result, contrary to parametric amplification, no synchronization is needed between propagating and modulating signals. In lossless media (or circuits), two main factors affect the amplification level of time-modulated defined media (circuit): the modulation depth and the loading impedance. Moreover, introducing loss to the system limits the amplification, as illustrated in section \ref{loss}. %It can be seen that a much weaker interaction happens between the harmonics $\omega_{1}$ and $\omega_{-2}$. This interaction may play a role in amplifying the main signal $\omega_s$ if the UC has the same electrical length (EL) at $\omega_{o}$ and $\omega_{1}$. This can be investigated in more complex unit cells. 
	Fig.~\ref{DDLC}(b) shows a close-up of the MBG region for different modulation depths. With a higher value of $M_D$, a wider MBG can be obtained. The phases at the edges of each band gap are written to be compared with the unit cell's electrical length (EL), which is discussed in the following paragraph. Fig.~\ref{DDLC}(c) shows the phase variation at the edges of the MPG with  $M_D$. In the case of no-modulation ($M_D=0$), the MBG edges match in phase, no MPG. However, as $M_D$ increases,  the MBG edges' phases deviate away, indicating an increase in the gap width. Moreover, the phase deviation is not symmetric at both edges because only part of the TL is a time-variant component, which makes the relation between varying the TMC and unit cell Bloch impedance ($Z_{Bloch}$) nonlinear. Based on edges' phase variation and analogy to mode coupling coefficient definition \cite{Hunter2001}, the harmonics mixing coefficient ($\kappa$) is plotted in Fig.~\ref{DDLC}(c) following   
	\begin{equation}
	\kappa=\frac{\left({\rm Phase}_{edge1}\right)^2-\left({\rm Phase}_{edge2}\right)^2}{\left({\rm Phase}_{edge1}\right)^2+\left({\rm Phase}_{edge2}\right)^2}
	\label{eq:coup}
	\end{equation}
	Within the MPG, only complex frequencies associated with fundamental $\omega_s$ and the harmonic $\omega_{-1}$ exist and are plotted in Fig.~\ref{DDLC}(d-g) by using the following equation instead of (\ref{eq:om}) in solving the eigenvalue problem.  
	
	\begin{equation}
	\Omega=\omega_n-i\sigma_n
	\label{eq:omc}
	\end{equation} 
	
	As shown Fig.~\ref{DDLC}(d), imaginary DD is plotted considering harmonics ($\omega_{0,-1}$) at different values of $M_D$ with a fixed real frequency $F_s=0.5$ GHz and varying the imaginary part. The positive imaginary frequency causes the signal growth, while the negative imaginary frequency is responsible for the signal decay. As shown, there is a direct relation between imaginary part values, the MPG width, and the value of $M_D$. Within MPG, as  $M_D$ increases, the strength of the harmonic ($\omega_{-1}$) and the imaginary frequency increase. Consequently, at a sufficient value of $M_D$, signal growth starts. This sufficient value of $M_D$ depends on the unit cells, their number, the matching at the terminals (load and source), and $F_m$. Both real and imaginary ($F_s=0.5$ GHz real part) DDs are plotted in Fig.~\ref{DDLC}(e) at $M_D=0.66$ to show that only within the MPG imaginary frequencies exist. Deviating from the frequency real part value of $F_s=0.5$ GHz, the imaginary part gradually vanishes due to the existence of real frequency solutions of the eigenvalue problem, as shown in Fig.~\ref{DDLC}(f). In other words, time modulation causes the creation of MPGs at only frequency ratio $F_m= 2F_{s}$ \cite{Galiffi2023}. Within the MPGs, only an imaginary frequency solution exists. Now, focusing on the unit cell electrical length at $C_o$ ($0.81$ rad), imaginary part variation of the fundamental $\omega_s$ and the harmonic $\omega_{-1}$, at fixed real part of $F_s=0.5$ GHz value, is plotted in Fig.~\ref{DDLC}(g) with $M_D$. As shown, the harmonic $\omega_{-1}$ imaginary part is directly proportional to the value of $M_D$.

	\begin{figure}[!t]
		\centering
		\includegraphics[width=.65\linewidth]{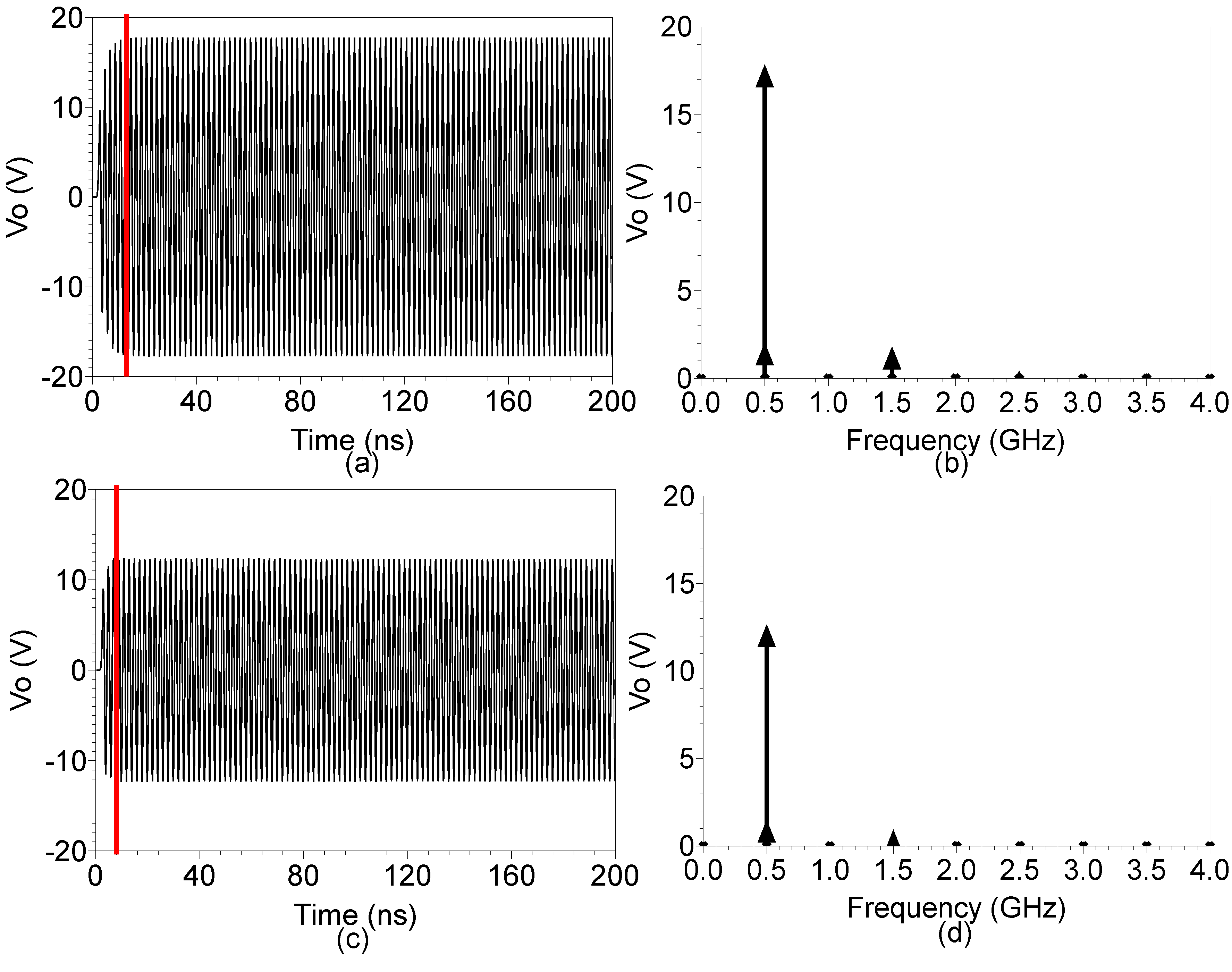}
		\caption{For 9 unit cells (Fig.~\ref{MODLE}(b)) with 10~V input peak (1 Watt(rms)), $L= 12.5$~nH, $C_o =5$~pF, $F_s =0.5$~GHz, $F_m =1$~GHz, loaded with a $50~\Omega$ load impedance. (a) Output voltage (TS) at TMC at $M_D =0.5$. (b) Output voltage (HB simulation) at TMC with $M_D =0.5$. (c) Output voltage (TS) at TMC with $M_D =0.33$. (d) Output voltage (HB simulation) at TMC with $M_D =0.33$.}
		\label{fig:lctr}
	\end{figure}

	\begin{figure}[!t]
		\centering
		\includegraphics[width=.65\linewidth]{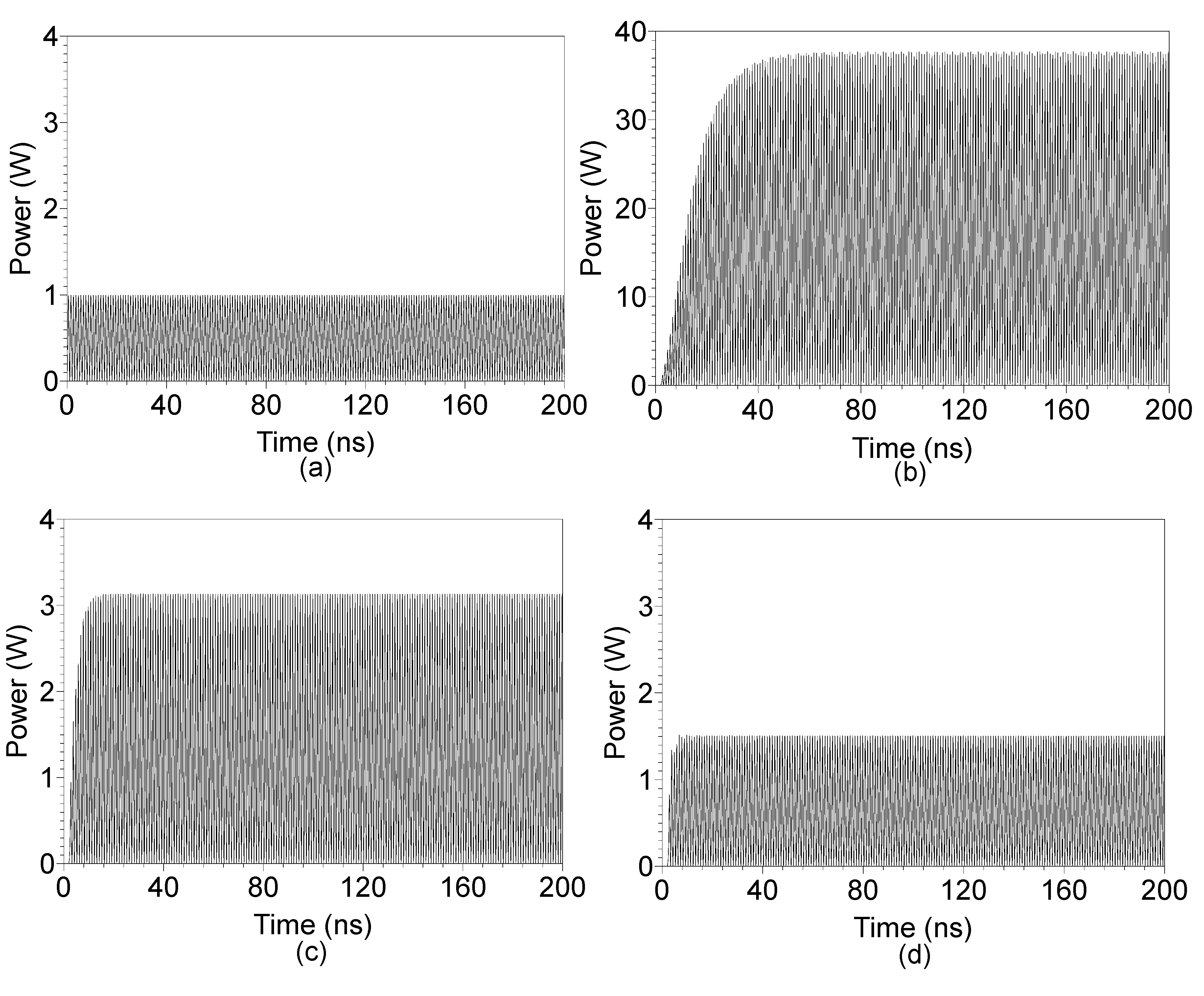}
		\caption{For 9 unit cells (Fig.~\ref{MODLE}(b)) with 10~V input peak (1 Watt(rms)), $L= 12.5 $~nH, $C_o =5$~pF, $F_s =0.5$~GHz, $F_m =1$~GHz, loaded with a $50~\Omega$ load impedance. (a) Input instantaneous power (rms) in the absence of modulation (TS). TS output power (rms) with TMC at (b) $M_D =0.66$, (c) $M_D =0.5$, and (d) $M_D =0.33$.}
		\label{fig:pow lc}
	\end{figure}

	%\begin{figure}[!t]
	%	\centering
	%	\includegraphics[width=.52\linewidth]{TLUC}
	%	\caption{UC of an ideal lossless TL loaded with a TMC. The TL has a length of $m\lambda_m$.}
	%	\label{MODLE}(c)
	%\end{figure}

	SP, TS, and HB simulations in ADS are utilized for further investigation. Fig.~\ref{lccir}(a) shows the phase of $S_{21}$ of the unit cell shown in Fig.~\ref{MODLE}(b) with $L= 12.5$~nH and $C= 5$~pF (static), done by S-parameter simulation. The unit cell has an absolute EL of $0.81$ radians at $F_s= 0.5$~GHz. This phase lies within the MBG illustrated in Fig.~\ref{DDLC}. In addition, to check the feasibility of (\ref{eq:N}), for a unit cell that has an absolute EL of $0.81$ radians, $\frac{\lambda_s}{d}$ gives $7.75$. So, choosing N=3 satisfies (\ref{eq:N}). On the other hand, it can be seen in Fig.~\ref{DDLC}(a) that harmonics $\omega_{\pm3}$ do not play any role at $F_s = 0.5$~GHz. A FBG occurs for these harmonics within the 0-1~GHz frequency range. By choosing N=2, the same DD of Fig.~\ref{DDLC} can be obtained. As a result, the correct value of N must satisfy (\ref{eq:N}), and it is also the lowest value above which the DD no longer changes within the frequency band of interest. Solutions to the eigenvalue problem do not come in order. This makes relating each curve to its corresponding harmonic tricky. Consequently, reducing the number of considered harmonics (N) to the lowest possible value decreases the complexity of the DD and helps to identify the harmonics' curves. However, six harmonics (N=3) are considered in the whole manuscript to account for any changes in the DD when the unit cell becomes more complex.

	For nine unit cells with TMCs at $M_D =0.66$ and $F_m =1$~GHz loaded with a $50~\Omega$ load impedance, Fig.~\ref{lccir}(b) shows the transient simulation with an input of 10~V (peak) matched source (1~W (rms)) at $F_s =0.5$~GHz. It is worth mentioning that 50~$\Omega$ matched voltage sources are utilized here and connected in series with an ideal 50~$\Omega$ isolator. The output power after the isolator is 1~W (rms), and the output voltage is 10~V peak. Within a 200~ns simulation period, the output voltage is amplified, reaching a 65~V peak level. For the same circuit, HB results are shown in Fig.~\ref{fig:lctr}(c-e). Figs.~\ref{fig:lctr}(d-e) are close-ups of the harmonics of interest. It can be seen that there are two amplified harmonics landed at 0.5~GHz ($n=0,-1$) and another two harmonics ($n=-2,1$) landed at 1.5~GHz. There is an agreement between the simulated amplification results obtained using HB and TS. The output amplified voltage from TS is almost equal to the summation of the two harmonics ($n=0,-1$) landed at 0.5~GHz obtained from HB. Usually, TS is more accurate than HB in highly nonlinear circuits. However, HB shows the output harmonics, which is very useful to verify the results depicted in Fig.~\ref{DDLC}. For lower values of $M_D$, Fig.~\ref{fig:lctr}(a-b) and (c-d) show the simulation results at $M_D=0.5$ and $M_D=0.33$, respectively. At $M_D=0.5$, the output voltage (TS, Fig.~\ref{fig:lctr}(a)) saturates after 12~ns at 18~V (peak), which is close to the summation of the two harmonics (HB, Fig.~\ref{fig:lctr}(b)) landed at 0.5~GHz. At $M_D=0.33$, the output voltage (TS, Fig.~\ref{fig:lctr}(c)) saturates after 8~ns at 13~V, which is exactly the summation of the two harmonics (HB, Fig.~\ref{fig:lctr}(d)) landed at 0.5~GHz. Fig.~\ref{fig:pow lc} shows the TS instantaneous output powers in different cases. Fig.~\ref{fig:pow lc}(a) shows the input power with a value of 1~W. The same value of input power will be used in the whole manuscript. When the TMC is utilized, the instantaneous output power is amplified. Increasing the modulation depth value causes higher output power values at the load. As shown in Fig.~\ref{fig:pow lc}(b-d), the output average power is saturated at 37~W, 3~W, and 1.4~W when modulating with $M_D=0.66$, $M_D=0.5$, and $M_D=0.33$, respectively. The simulated results confirm that power amplification is directly proportional to the modulation width $M_D$.

	\section{Signal Amplification in a lossless TL loaded  time modulated capacitor}
	In this section, the element (A) in Fig.~\ref{MODLE}(a) is replaced by a TL section with an electrical length of $2m\pi$ radian at the modulation frequency and $90~\Omega$ characteristic impedance ($Z_o$). Two conditions are considered for element (B) in Fig.~\ref{MODLE}(a): Infinite quality factor (Q) TMC and finite Q TMC. 
 \subsection{Infinite quality factor time modulated capacitor }
 
 \begin{figure*}[!t]
		\centering
		\includegraphics[width=1.03\linewidth]{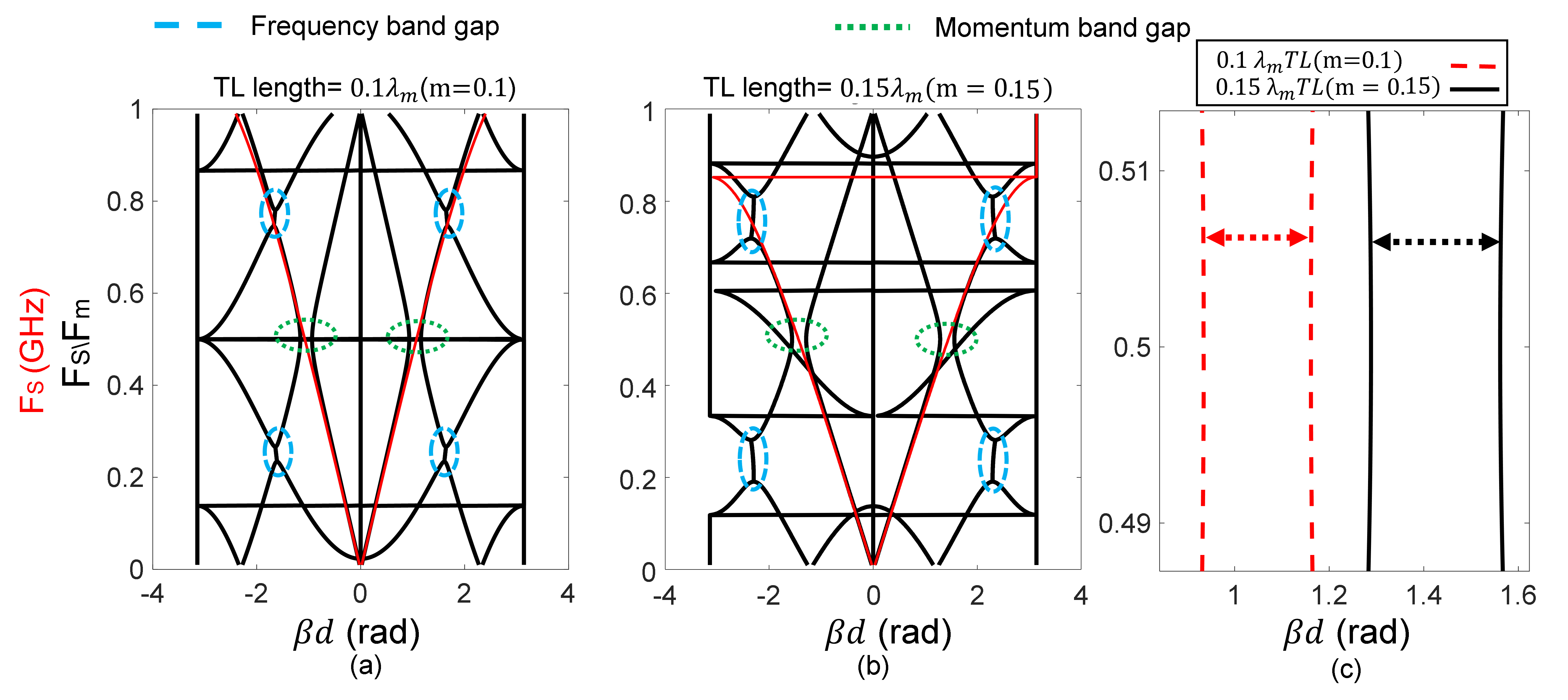}
		\caption{DD of the unit cell in Fig.~\ref{MODLE}(c) with $C_o =4 $~pF, $M_D=0.5$, $90~\Omega$ characteristic impedance ($Z_o$), and  TL sections with different lengths. (a) TL length $0.1 \lambda_m$ ($m=0.1$), and (b) TL length $0.15 \lambda_m$ ($m=0.15$). (c) Close-up look of the MBGs.}
		\label{fig:DDTL}
	\end{figure*}

 In this subsection,  element (B) will be an infinite Q TMC. The unit cell can be seen in Fig.~\ref{MODLE}(c), and different cases are investigated at various values of $m$. In order to plot the DD, (\ref{eq:dd}) and (\ref{eq:dd1}) are solved with the matrix $T_L$ in (\ref{eq:TT}) replaced by
	
	\begin{equation}
	T_{tl}=\left[\begin{matrix}\bar{\bar{A}}&\bar{\bar{B}}\\\bar{\bar{C}}&\bar{\bar{D}}\\\end{matrix}\right]
	\label{eq:TTL}
	\end{equation}
	where
	\begin{subequations}
		\begin{align}
		\bar{\bar{A}} & = \cos (2\times \pi \times m \times \frac{\bar{\bar{W}}}{\omega_M}) ) \\
		\bar{\bar{B}} & = j \times Z_o \times \sin (2\times \pi \times m \times \frac{\bar{\bar{W}}}{\omega_M}) ) \\
		\bar{\bar{C}} & = j \times Y_o \times \sin (2\times \pi \times m \times \frac{\bar{\bar{W}}}{\omega_M}) ) \\
		\bar{\bar{D}} & = \cos (2\times \pi \times m \times \frac{\bar{\bar{W}}}{\omega_M}) ) 
		\end{align}
	\end{subequations}

	\begin{figure}[!t]
		\centering
		\includegraphics[width=.65\linewidth]{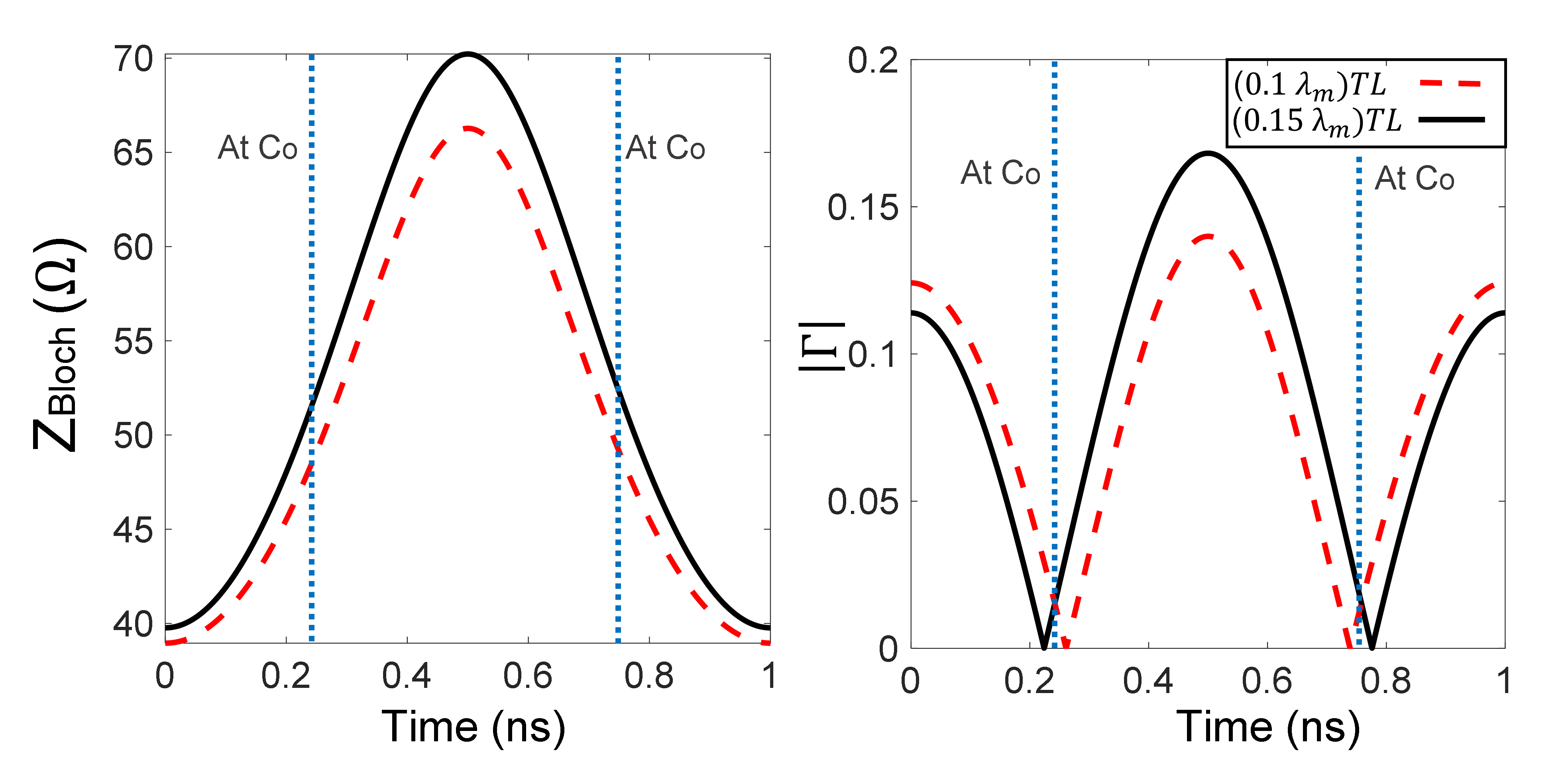}
		\caption{For unit cell in Fig.~\ref{MODLE}(c) with $C_o =4 $~pF, $M_D=0.6$, $90~\Omega$ characteristic impedance ($Z_o$), TL section lengths $0.1 \lambda_m$ and $0.15 \lambda_m$, at the fundamental frequency (0.5~GHz). (a) Variation in Bloch impedances across time for one complete cycle. (b) The absolute value of the reflection coefficient ($\lvert \Gamma \rvert$) within one complete cycle considering $50~\Omega$ load (source).}
		\label{fig:Zbloch}
	\end{figure}

	Adding TL to the unit cell in Fig.~\ref{MODLE}(a) as element (A) allows us to investigate the effect of varying the electrical length of the TL on the amplification. Changing the electrical length of the time-modulated unit cell will allow the control of the location of the FBG (space) in the DD of one of the prominent harmonics ($\omega_{0,-1}$) relative to the MBG (time). The DD (N=3) of the unit cell shown in Fig.~\ref{MODLE}(c) is plotted in Fig.~\ref{fig:DDTL} for different lengths of the TL. $90~\Omega$ $Z_o$ TL sections are utilized to match the unit cell at nominal capacitance $C_o$ to a $50~\Omega$ load. Red curves in Fig.~\ref{fig:DDTL} represent the dispersion curve of the unit cell without any modulation. In the no-modulation case, it can be seen that varying the TL length changes the position of the FBG that starts from and ends by phase $\pm \pi$ \cite{pozar2011microwave}. With time modulation, the MBG appears at $F_s= 0.5$~GHz, considering $F_m= 1$~GHz, green circles note the MBGs. Moreover, due to the interaction between harmonics, FBGs are created at phase values different than $\pm \pi$. The FBGs are noted in blue circles. We are interested in FBGs occurring by one of the prominent harmonics ($\omega_{0,-1}$). As shown in Fig.~\ref{fig:DDTL}, as the FBG becomes closer to the MBG, the MBG becomes wider, and the slope of the dispersion curves around the MBG increases. Wider MBG and slopes indicate higher amplification values and broader response around $F_s= 0.5$~GHz.
 
 \begin{figure}[!t]
		\centering
		\includegraphics[width=1.03\linewidth]{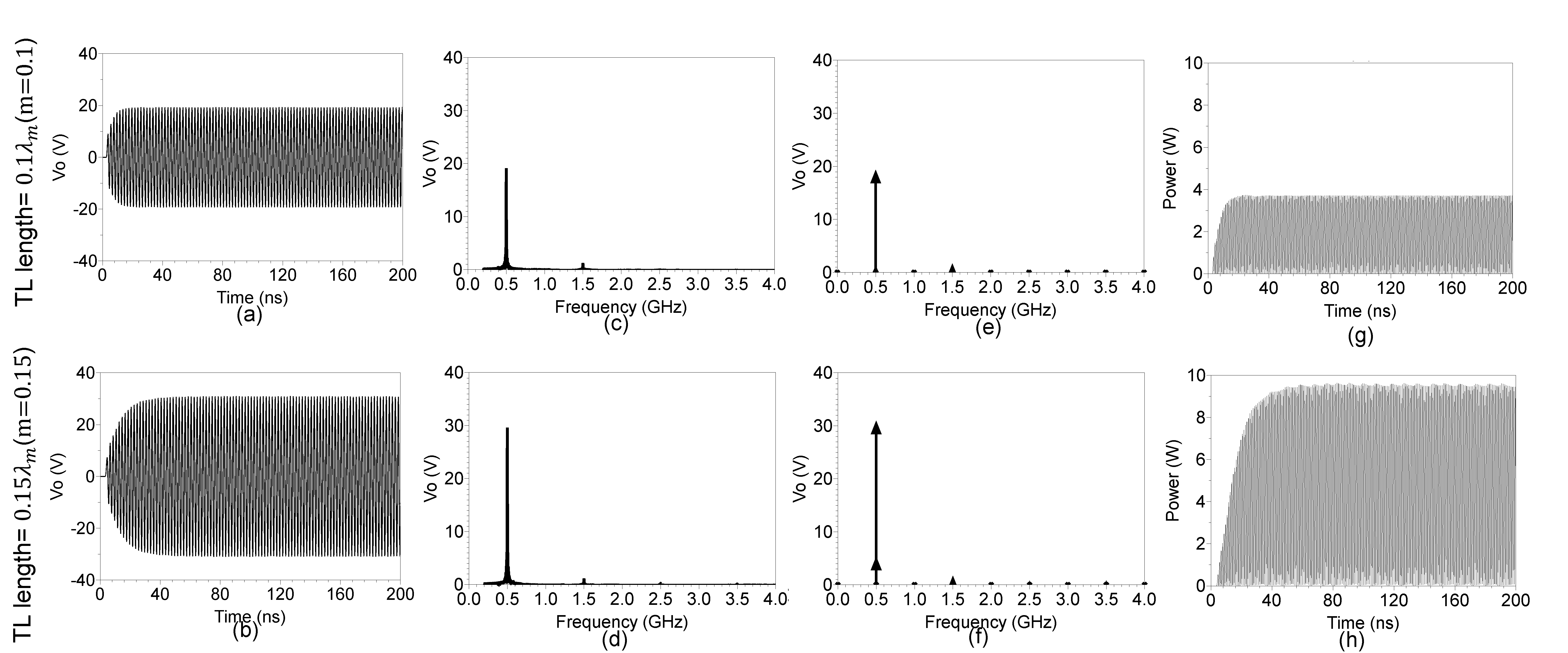}
		\caption{For 9 unit cells (Fig.~\ref{MODLE}(c)) with 10~V input peak (1 Watt (rms)), $C_o =4$~pF $F_s =0.5$~GHz, $F_m =1$~GHz, $M_D =0.6$, and the TL with $90~\Omega$ characteristic impedance ($Z_o$) and different lengths loaded with a $50 \Omega$ load impedance. For $0.1 \lambda_m$ ($m=0.1$) TL: (a) output voltage (TS), (c) Fourier transform (FT) of the output voltage (TS), (e) output voltage HB, and (g) power output (TS). For $0.15 \lambda_m$ ($m=0.15$) TL: (b) output voltage (TS), (d) Fourier transform (FT) of the output voltage (TS), (f) output voltage HB, and (h) power output (TS).}
		\label{fig:AMPTL}
	\end{figure}

    To further investigate the effect of TL length on the MBG width, the variation across time of the unit cell Bloch impedances are plotted in Fig.~\ref{fig:Zbloch} for the fundamental frequency of 0.5~GHz. As in \cite{pozar2011microwave}, Bloch impedance can be plotted using
    \begin{equation}
	Z_{Bloch}^\pm=\frac{-2B}{A-D\mp\sqrt{{(A+D)}^2-4}}
	\label{eq:ZBLOCH}
    \end{equation}  
	where A, B, C, and D are the ABCD matrix elements of the unit cell at the fundamental frequency ($\omega_s$). The $\pm$ solutions correspond to $Z_{Bloch}$ for forward and backward traveling waves, respectively.  Fig.~\ref{fig:Zbloch}(a) shows the Bloch impedance variation for the unit cell of TL lengths $0.1 \lambda_m$ and  $0.2 \lambda_m$ sections within one complete cycle. It can be seen that a unit cell of $0.2 \lambda_m$ TL section has a slightly wider impedance variation range of 41-70~$\Omega$ compared to the unit cell of $0.1 \lambda_m$ TL section, which shows  40-66~$\Omega$ impedance variation range. This agrees with the results shown in Fig.~\ref{fig:DDTL}; a unit cell of $0.15 \lambda_m$ TL section has a wider MBG than the unit cell of $0.1 \lambda_m$  section. As $ M_D$ increases, a wider range of impedances and MBGs are also expected. In addition, the difference between unit cells of $0.1 \lambda_m$ and $0.15 \lambda_m$ TL sections in the MBG and impedance variation range becomes more apparent. Fig.~\ref{fig:Zbloch}(b) shows the absolute value of the reflection coefficient ($\lvert \Gamma \rvert$) within one complete cycle considering a $50~\Omega$ load (source), which is computed using
	\begin{equation}
	\Gamma_{load\ (source)}=\frac{Z_{load\ (source)}-Z_{Bloch}}{Z_{load\ (source)}+Z_{Bloch}}
	\label{eq:gama}
	\end{equation}

	As shown in Fig.~\ref{fig:Zbloch}(b), the values of $\lvert \Gamma \rvert$ are almost the same except at the half of the cycle, where $\lvert \Gamma \rvert$ is higher for $0.2 \lambda_m$ TL unit cell compared to $0.1 \lambda_m$ TL unit cell.  Increasing $\lvert \Gamma \rvert$ could enhance the amplification for $0.2 \lambda_m$ TL unit cell because the reflected signal can be added constructively to the main signal and re-amplified \cite{Galiffi2023}. However, increasing the mismatch at the terminals can cause instability and oscillation.

	TS are performed for 9 unit cells (Fig.~\ref{MODLE}(c)) with $50~\Omega$ matched source suppling 10~V peak (1W ), $C_o =4$~pF, $F_s =0.5$~GHz, $F_m =1$~GHz, $M_D =0.6$, and a TL with $90~\Omega$ characteristic impedance ($Z_o$) and different lengths loaded with a $50~\Omega$ load impedance. Two TL lengths are chosen, $0.1 \lambda_m$ and $0.15 \lambda_m$, to be almost matched at $C_o$. The results are shown in Fig.~\ref{fig:AMPTL}, where signal amplification is observed. The unit cell consists of ideal elements, and introducing loss will make the circuit saturate at lower values and affect the system's responsivity to the modulation. Figs. \ref{fig:AMPTL}(a,b) show the TS at TL lengths of $0.1 \lambda_m$ and $0.15 \lambda_m$, respectively. Figs. \ref{fig:AMPTL}(c,d) show Fourier transforms of the results obtained in Figs. \ref{fig:AMPTL}(a,b), receptively. As shown, as the TL length increases, higher amplification values are obtained; the output voltage is amplified to 18~V and 28~V using TL lengths of $0.1 \lambda_m$ and $0.15 \lambda_m$, respectively. Figs.~\ref{fig:AMPTL}(e,f) show the HB simulation results, which agree with the TS results shown in Figs.~\ref{fig:AMPTL}(a,b). The input power is the same as shown in Fig.~\ref{fig:pow lc}(a). The instantaneous output power at TL lengths $0.1 \lambda_m$ and $0.15 \lambda_m$ with modulation are shown in Figs.~\ref{fig:AMPTL}(g,h), respectively.  The instantaneous output power is amplified to 3.5~W and 9.2~W using TL lengths of $0.1 \lambda_m$ and $0.15 \lambda_m$, respectively. The obtained simulation results match the analysis done using the DD and Bloch impedance. Compared to the unit cell of  $0.1 \lambda_m$ TL section, the unit cell of $0.15 \lambda_m$ TL section has a wider MBG and variation range of Bloch impedances. Consequently, as discussed, the unit cell of $0.15 \lambda_m$ TL section has higher voltage and power amplification values.

	%\begin{figure}[!t]
	%	\centering
	%	\includegraphics[width=.52\linewidth]{TLUCR}
	%	\caption{UC of an ideal lossless TL loaded with a limited quality factor temporally modulated capacitor. The TL has a length of $0.15 \lambda_m$ (m=0.15) and $90~\Omega$ characteristic impedance.}
	%	\label{MODLE}(d)
	%\end{figure} 

	 \subsection{Finite quality factor time modulated capacitor }
  \label{loss}
	In this subsection, element (A) in Fig.~\ref{MODLE}(a) is a TL with length $0.15 \lambda_m$ (m=0.15) and $90~\Omega$ characteristic impedance ($Z_o$). Element (B) will be a TMC with series resistance (Rc). The introduced resistance Rc limits the quality factor of the TMC. The unit cell is illustrated in Fig. \ref{MODLE}(d) and will be studied at different values of Rc. 
	
	\begin{figure*}[!t]
		\centering
		\includegraphics[width=1\linewidth]{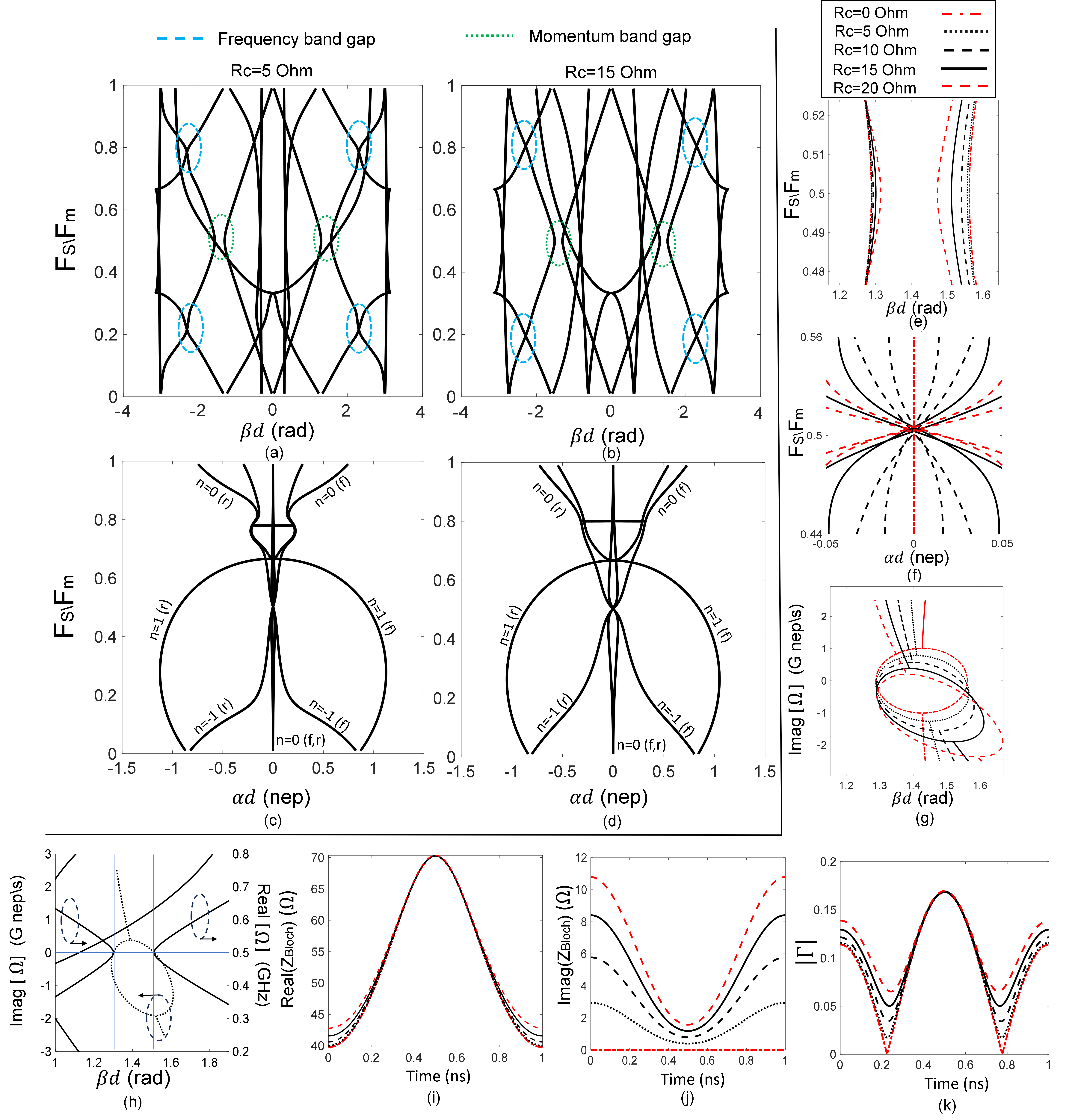}
		\caption{ For the unit cell in Fig.~\ref{MODLE}(d) with $C_o =4$~pF, $M_D=0.6$, and a TL with $90~\Omega$ characteristic impedance ($Z_o$) and length of $0.15 \lambda_m$ (m=0.15). DD at (a) Rc~=~5~$\Omega$ and
			(b) Rc~=~15~$\Omega$. Attenuation variation versus $F_s/F_m$ at (c) Rc~=~5~$\Omega$ and
			(d) Rc~=~15~$\Omega$. For various values of Rc (0, 5, 10, 15, 20)~$\Omega$, (e) Close-up look of the MBG, (f) The imaginary part of complex frequency variation within the MBG at $F_s/F_m=0.5$ for harmonics $\omega_o$ and $\omega_{-1}$ and (g)  Close-up look of attenuation at the MBG. (h) For Rc~=~15~$\Omega$, the Unit cell electrical length variation with the real and imaginary parts (real is fixed at $F_s/F_m=0.5$ for imaginary part variation) of the complex frequency. (i) The real part of Bloch impedance variation within one complete cycle. (j) The imaginary part of the Bloch impedance variation within one complete cycle. (k) The absolute value of the reflection coefficient ($\lvert \Gamma \rvert$) within one complete cycle considering $50~\Omega$ load (source).  }
		\label{fig:DDTLR}
	\end{figure*}

	To plot the DD, the transfer matrix of the shunt element (B) should be computed. The resistance matrix is given by
	\begin{equation}
	\bar{\bar{Z_{Rc}}}=Rc\times \bar{\bar{Ones}} 
	\label{eq:Rc}
	\end{equation}
	where Rc is the value of the resistance. The impedance matrix of the shunt branch B can be calculated using 
	\begin{equation}
	\bar{\bar{Z_{B}}}= \bar{\bar{Z_{Rc}}} + {\bar{\bar{Y_C}}}^{-1} 
	\label{eq:ZB}
	\end{equation}
	where $\bar{\bar{Y_C}}$ is given in (\ref{eq:YC}). The transfer matrix of shunt branch B is given by
	\begin{equation}
	T_B=\left[\begin{matrix}\bar{\bar{Ones}}&\bar{\bar{Zeros}}\\\ {\bar{\bar{Z_{B}}}}^{-1}&\bar{\bar{Ones}}\\\end{matrix}\right]
	\label{T_B}
	\end{equation}	 
	The total transfer matrix of the unit cell shown in Fig.~\ref{MODLE}(d) can be obtained using
	\begin{equation}
	T_{Rc}= T_{tl} \times T_B \times T_{tl}
	\label{eq:TT1}
	\end{equation}
	where $T_{tl}$ is given in (\ref{eq:TTL}) with m = 0.15.
	
	The DDs are plotted in Fig.~\ref{fig:DDTLR} utilizing (\ref{eq:TT1}) and (\ref{eq:dd1}). In Figs. \ref{fig:DDTLR}(a)(b), the DD is plotted at Rc values of 5 and 15~$\Omega$, respectively. Compared to Fig.~\ref{fig:DDTLR}(b), weak interaction between harmonics can be observed in the areas of the FBGs, circled in blue, and the MBGs, circled in green. As the value of Rc increases, the FBG and MBG areas shrink. Moreover, the attenuation constant variations with frequency associated with the cases shown in  Figs. \ref{fig:DDTLR}(a)(b) are shown in  Figs. \ref{fig:DDTLR}(d)(c), respectively. Focusing on the prominent harmonics ($\omega_{0,-1}$), the attenuation constant increases at all frequencies as the Rc value increases except for $F_s = 2F_m$; attenuation is always zero as long as the MPG exists. To clearly show the effect of increasing the value of Rc on the MBG and attenuation, a close-up look of the MBG and attenuation at $F_s = 2F_m$   at different values of Rc are plotted in Fig.~\ref{fig:DDTLR}(e),(f), respectively. As shown in Fig.~\ref{fig:DDTLR}(e), the MPG shrinking is obvious as an effect of increasing the value of Rc. On the other hand, as shown in Fig.~\ref{fig:DDTLR}(f), despite hitting the zero value of attenuation at $F_s = 2F_m$ regardless of the value of Rc, attenuation at other frequencies increases with the increase of Rc. It is worth mentioning that if Rc reaches a value that closes the MPG, attenuation will not be zero anymore at $F_s = 2F_m$. 
	In Fig.~\ref{fig:DDTLR}(g), imaginary DD is plotted at different values of Rc considering harmonics ($\omega_{0,-1}$) with a fixed real frequency $F_s=0.5$ GHz. As Rc increases, the imaginary frequency moves more to the negative region, explaining the expected drop in amplification levels. In Fig.~\ref{fig:DDTLR}(h), real and imaginary (real at  $F_s=0.5$) DDs are plotted at Rc= 15$\Omega$. The MBG width is mainly defined by the phases at which the positive imaginary frequency exists. In this case (Rc= 15$\Omega$) and within the MBG, the imaginary frequency takes the values confined between the vertical blue dotted lines because, outside this region, real solutions to the eigenvalue problem exist.  
	\begin{figure}[!t]
		\centering
		\includegraphics[width=1\linewidth]{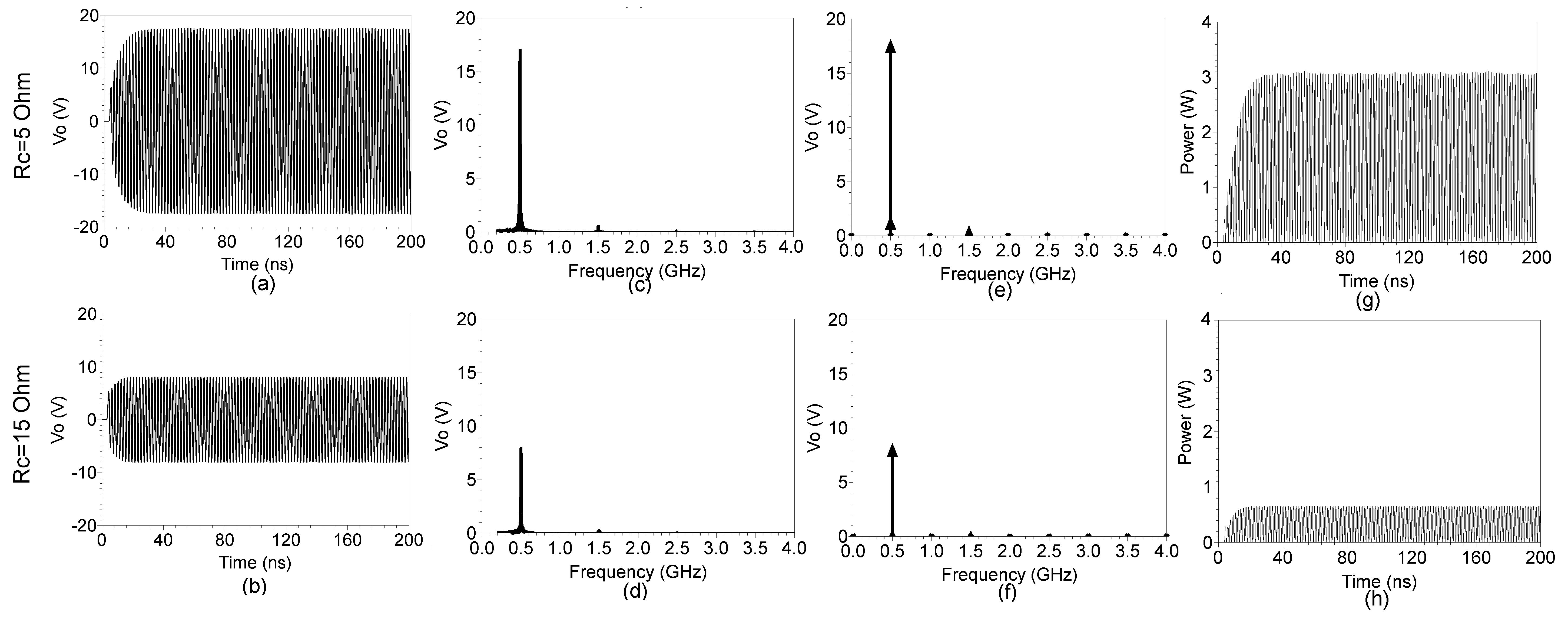}
		\caption{For 9 unit cells (Fig.~\ref{MODLE}(d)), with 10~V input peak (1 Watt(rms)), $C_o =4$~pF, $F_s =0.5$~GHz, $F_m =1$~GHz, $M_D =0.5$, a TL with $90~\Omega$ characteristic impedance ($Z_o$) and length of $0.15 \lambda_m$ loaded with a $50~\Omega$ load impedance. (a-b) Voltage output (TS) at Rc~=~5~$\Omega$ and 15~$\Omega$, respectively. (c-d) Fourier transforms for TS results at Rc~=~5~$\Omega$ and 15~$\Omega$, respectively. (e-f) Voltage output (HB) at Rc~=~5~$\Omega$ and 15~$\Omega$, respectively. (g-h) Power output (TS) at Rc~=~5~$\Omega$ and 15~$\Omega$, respectively.}
		\label{fig:AMPTLR}
	\end{figure}
	
	Figs.~\ref{fig:DDTLR}(i-k) show the effect of Rc on the Bloch impedance ($Z_{Bloch}$) and $\lvert \Gamma \rvert$ at the terminals. The variation range of the real part of $Z_{Bloch}$ shrinks as Rc increases (Fig.~\ref{fig:DDTLR}(i)). Moreover, the imaginary part of $Z_{Bloch}$ increases which reflects the loss increase within the unit cell (Fig.~\ref{fig:DDTLR}(j)). Moreover, the circuit gradually loses matching at the $50~\Omega$ terminals at the quarter cycle (nominal $C_o$) as Rc increases (Fig.~\ref{fig:DDTLR}(k)). Consequently, it can be concluded that the value of Rc is inversely proportional to the signal amplification gain.

	To confirm the observed weak interaction between harmonics and its relation to Rc, TS is performed with 9 unit cells (Fig.~\ref{MODLE}(d)) with a $50~\Omega$ matched source supplying 10~V peak (1~W ), $C_o =4$~pF, $F_s =0.5$~GHz, $F_m =1$~GHz, $M_D =0.6$, a TL with $50~\Omega$ characteristic impedance ($Z_o$) and length of $0.15 \lambda_m$ loaded with a $50~\Omega$ load impedance. Two values of Rc are considered, 5~$\Omega$ and 15~$\Omega$, and the results are plotted in Fig.~\ref{fig:AMPTLR}. For Rc~=~5~$\Omega$, the output voltage saturates at the peak of 17~V (Fig.~\ref{fig:AMPTLR}(a)(c)(e)), and the instantaneous output power saturates at 3~W (Fig. \ref{fig:AMPTLR}(g)). Increasing the value of Rc to 15~$\Omega$ causes a drop in the output voltage to 7~V peak value (Fig. \ref{fig:AMPTLR}(b)(d)(f)), and in the instantaneous output power to 0.6~W (Fig. \ref{fig:AMPTLR}(h)). As shown, when Rc reaches 15~$\Omega$, there is no amplification.
	 
	\section{Conclusion}
	
	Signal amplification in a TL with time-modulated $Z_o$ is investigated by studying the eigenvalue problem and confirmed by circuit modeling. Three models are considered: a lossless L-C TL lumped model with shunt TMC, a TL loaded with shunt infinite Q TMC, and a TL loaded with shunt finite Q TMC to study the loss effect. The eigenvalue problem is discussed in detail, and the real and imaginary DDs are plotted for each model. While modulation, the Bloch impedance of the unit cell and the reflection coefficient at the terminals are plotted and discussed.  SP, TS, and HB simulations are performed, and the results are consistent with the DD, Bloch impedance variation of the unit cell and reflection coefficient variation at the terminals. The third model discusses the loss effect by plotting the attenuation variation with frequency and real and imaginary DDs. In a lossy time-modulated media, if the loss is low enough to allow the MBG to be created at $F_s = 2F_m$, attenuation will be zero at $F_s = 2F_m$. However, the drop in amplification level happens because of the movement of imaginary frequency within the MBG to the negative region, causing the signal decay.   
\section*{Acknowledgements}

This work has been completed under a research agreement between Purdue University and The American University in Cairo.
 
\

\bibliography{bibfile1}% Produces the bibliography via BibTeX.

\end{document}